\documentclass[aps,showpacs,twocolumn,prl]{revtex4}
\usepackage{mathrsfs}
\usepackage{amssymb}
\usepackage{amsmath}
\usepackage{bm}
\usepackage{graphics}
\usepackage{natbib}

\begin{document}
\title{Enhancement of the thermoelectric effect due to the Majorana zero modes coupled to one quantum-dot system}
\author{Xiao-Qi Wang$^1$}
\author{Shu-Feng Zhang$^2$}
\author{Yu Han$^3$}
\author{Guang-Yu Yi$^1$}
\author{Wei-Jiang Gong$^1$}\email{gwj@mail.neu.edu.cn}
\affiliation{1. College of Sciences, Northeastern University, Shenyang 110819, China\\
2. Department of Physics, University of Jinan, Jinan 250022, China\\
3. Department of Physics, Liaoning University, Shenyang 110036,
China}
\date{\today}

\begin{abstract}
By considering Majorana zero modes to laterally couple to the quantum dot, we evaluate the thermoelectric effect in one single-dot system. The calculation results show that if one Majorana zero mode couples to the dot, the thermoelectric effect will exhibit its change, but the thermoelectric efficiency cannot be enhanced apparently. However, the thermoelectric effect can be efficiently strengthened when two Majorana zero modes are introduced simultaneously. We believe that the findings in this work provide an alternative method for the detection of Majorana bound state.
\end{abstract}
 \maketitle

\bigskip

\section{Introduction}
The topic of Majorana bound state (MBS) has attracted much attention, due to its
fundamental physics interest and potential applications. Many groups proposed ways to realize MBSs, such as in a vortex core in a p-wave
superconductor\cite{MBS1,Read,FuL,Sato,Sau,Alicea} or superfluid\cite{Kopnin,Sarma}. Then, it has been reported that they can be realized at the ends of a one-dimensional p-wave superconductor for which the proposed system is a semiconductor nanowire with Rashba spin-orbit interaction to which both a magnetic field and proximity-induced s-wave pairing are added\cite{Kitaev,Sau2,Oreg,Wu0}. Following these works, researchers suggested various schemes to detect the MBSs, including noise measurements\cite{Noise}, resonant Andreev reflection by a scanning tunneling miscroscope\cite{resonantAndreev}, the zero-bias conductance peak\cite{ZBP}, and the fractional Josephson effect\cite{frac-Jose,frac-Jose1,frac-Jose2}. Recent reports have further verified the existence of MBSs in solids\cite{Jia1,Jia2}.
\par
The MBS-detection suggestions indicate that the properties of MBSs can be clarified by embedding them in the mesoscopic circuit and investigating the transport behaviors. This conclusion should be ascertained, especially when the MBS couples to the regular fermion bound state in the quantum dot (QD). The reason is that the interplay between these two bound states is able to induce a variety of interesting phenomena.
For instance, in the T-shaped double QDs coupled to the MBS, the Andreev conductance spectrum presents a well-defined insulating
band in the low-bias region\cite{gong}. In a MBS-embedded Fano setup, the Fano effect becomes more complicated and determined by the structural parameters\cite{Fano1,Fano2}. When the QD eneters the Kondo regime, some important phenomena are induced due to the interaction between the Kondo effect and the Andreev reflection \cite{Andreevkondo1,Andreevkondo2,Andreevkondo3,Mkondo1,Mkondo2,Mkondo3}. It has been shown that in the structure where the MBS couples to the metal via one Kondo QD, the system flows to a new fixed point controlled by the Majorana-induced
coupling in addition
to the Kondo fixed point, which is characterized by the correlations between the QD and the fermion parity of the TS and metal\cite{Andreevkondo3}. On the other hand, if the MBS couples to one QD in one closed circuit, the zero-bias conductance value is halved, independent of the geometries the QD systems\cite{Liude}. Similar a result holds even when the QD is in the Kondo
regime\cite{LeeM}. Also, the MBS is characterized by its nonlocal nature, which can result in electron teleportation by means of nonlocal phase-coherent electron transfer through tunneling in and out of a pair of MBSs. Tunneling experiment which can uniquely identify the nonlocal electron tunneling assisted by MBSs has been proposed\cite{Liuj}. Recently, the emergence of MBSs in a hybrid InAs nanowire with epitaxial Al has been experimentally demonstrated. At the end of the nanowire QD was used as a
spectrometer\cite{NN}.
\par
In addition to the conventional quantum transport properties, the MBS-driven thermoelectric effect has attracted much attention. The motivation originates from the fact that its-existed structure also possesses the $\delta$-like density of states, which benefits for the enhancement of thermoelectric efficiency\cite{Mahan2,Mahan3}. Firstly, the thermoelectric effect in the metal/QD/MBS structure has been investigated. The authors found that the resulting gate-dependent Seebeck coefficient provides a new way to evidence
the existence of MBSs, which can be combined with conventional tunnel
spectroscopy in the same setup\cite{Thermal1}. Next, some groups discussed the thermoelectric properties of nanowires hosting MBSs, by coupling the nanowire to two normal metallic leads. It has been observed that the thermopower always vanishes regardless of the value of the Majorana hybridization. However, this situation changes drastically if one QD is inserted, and the thermoelectric effect and be efficiently modulated by the QD-MBS coupling\cite{Thermal2}. Besides, the MBS takes nontrivial effect to the violation of the Wiedemann-Franz law\cite{Thermal3}. Moreover, it has been shown that the interaction between the MBS and the Kondo QD enables to induce some interesting thermoelectric phenomena, and the thermoelectric quantities exhibit new properties\cite{Thermal4}.
\par
The previous works indeed show that the MBS-driven thermoelectric effect is very important and worthy of further investigation. Motivated by such a fact, in the present work we would like to investigate the thermoelectric effect in one QD system, by considering MBSs to couple to the QD. As a typical case, we are only interested in the presence of Majorana zero modes (MZMs). Via calculation, we find that if two MZMs couple to the QD simultaneously, thermoelectric effect can be enhanced to a great extent. We then believe that these results can be helpful in further understanding the quantum transport property of the MZM. Also, they provide an alternative method for the MBS detection.
\begin{figure}
\centering \scalebox{0.33}{\includegraphics{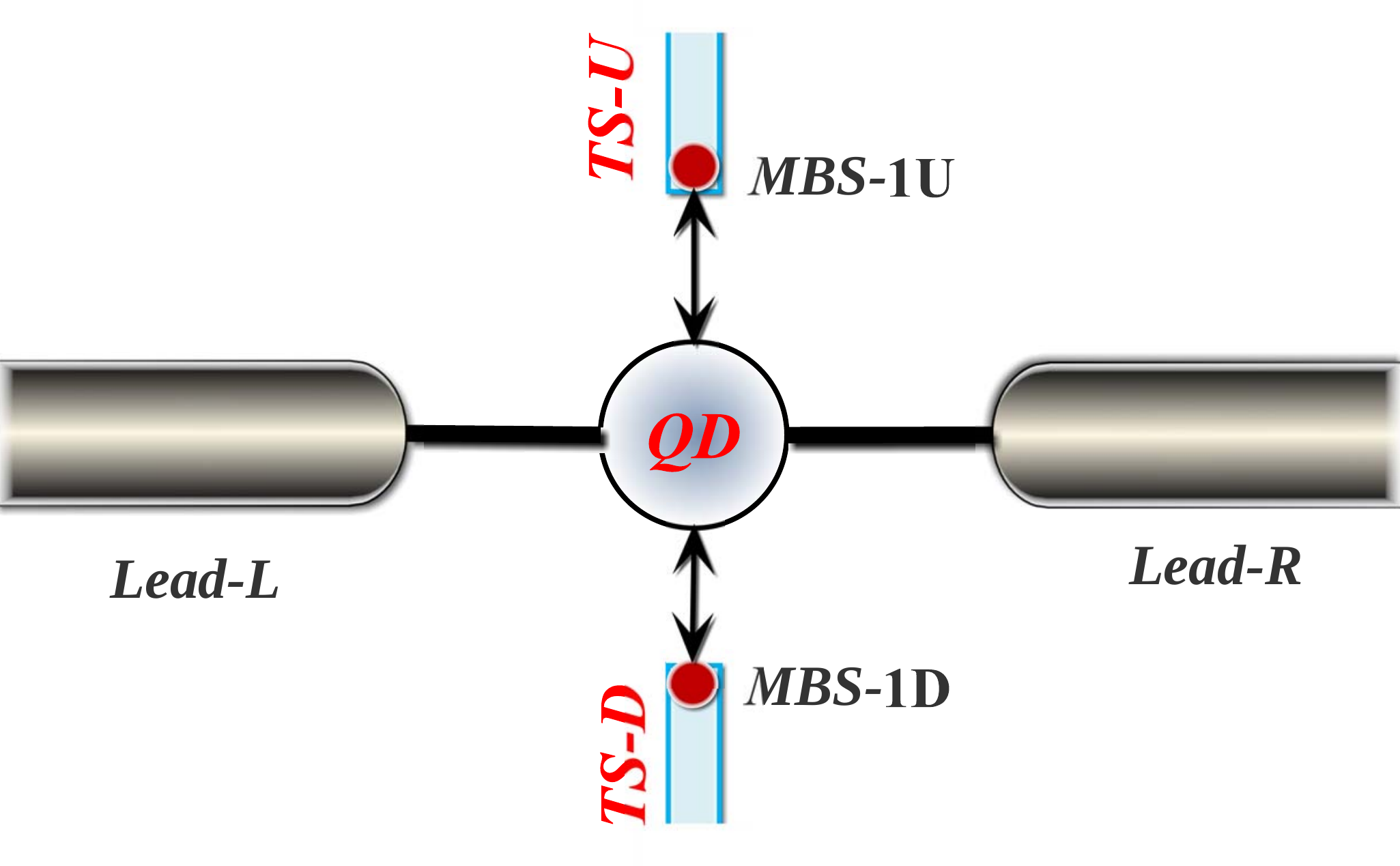}} \caption{
Schematic of one single-QD structure coupled to two topological-superconductor (TS) nanowires. The two nanowires, i.e., TS-$U$ and TS-$D$, are assumed to contribute MBSs at their ends. In this structure, only one MBS in each TS nanowire couples to the QD.     \label{structure}}
\end{figure}
\section{The theoretical model\label{theory}}
The Hamiltonian for our considered structure can be written as
$H=H_{0}+H_M+H_{T}$. The first term is the Hamiltonian for the two normal metallic leads, QD, and their couplings. It takes the form as
\begin{eqnarray}
H_{0}=\sum_{\alpha k }\varepsilon _{\alpha k}c_{\alpha k }^\dag
c_{\alpha k }+\varepsilon_d d^\dag
d+\sum_{\alpha k}V_{\alpha} c^\dag_{\alpha
k}d+h.c..\label{2}
\end{eqnarray}
$c_{\alpha k }^\dag$ $( c_{\alpha k })$ is an operator to create
(annihilate) an electron of the continuous state $|k \rangle$ in
lead-$\alpha$ ($\alpha\in L,R$), and $d^\dag$ ($d$) is the creation (annihilation) operator in the QD. $\varepsilon_{\alpha k}$ and $\varepsilon_d$ are the corresponding levels. Here the Coulomb interaction in the QD is ignored, since we are mainly interested in the interplay between the electronic bound state and MBSs. Next, the low-energy effective
Hamiltonian for $H_M$ (i.e., the MBS Hamiltonian) reads
\begin{equation}
H_{M}=i\sum_{n=U,D}t_{n}\eta_{1n}\eta_{2n},
\end{equation}
where $\eta_{jn}$ ($j=1,2$ and $n=U,D$) is the self-Hermitian operator for the $j$-th MBS
in the $n$-th topological superconducting nanowire with $\eta_{jn}=\eta_{jn}^\dag$. The last term of $H$
denotes the coupling between the QD and MBSs. In our
structure, we focus on the case where MBS-$({1n})$ is coupled to the
QD, so $H_T$ can be directly written as
\begin{eqnarray}
H_{T}&=&\sum_{n}(\lambda_{n} d^\dag-\lambda^*_{n}
d)\eta_{1n}.\label{TT}
\end{eqnarray}
$\lambda_n$ is the coupling coefficient between $\eta_{1n}$ and the QD.
\par

In such a structure, the electric and heat current can be defined as
a change in the number of electrons and the total energy per unit
time in lead-$L$, respectively. With the help of the nonequilibrium Green function technique, the
electric and heat currents can be expressed as\cite{Meir1,Meir2,Kim1,Kim2,themalgong}
\begin{eqnarray}
J^L_e&=&{e\over h}\int d\omega
\tau(\omega)[f_L(\omega)-f_R(\omega)],\notag\\
J^L_Q&=&{1\over h} \int d\omega
(\omega-\mu_L)\tau(\omega)[f_L(\omega)-f_R(\omega)].
\end{eqnarray}
$f_\alpha(\omega)=[\exp{\omega-\mu_\alpha\over k_BT_\alpha}+1]^{-1}$
is the Fermi distribution function of lead-$\alpha$ when each lead
is in thermal equilibrium at temperature $T_\alpha$.
$\mu_{\alpha}={eV_\alpha}$ is the chemical potential shift due to
the applied source-drain bias voltage $V_\alpha$. The transmission
spectral function $\tau(\omega)$ is given by the following
expression\cite{Meir1}
\begin{equation}
\tau(\omega)=-\Gamma {\mathrm{Im}}G^r_{dd}.\label{transmission}
\end{equation}
$\Gamma$ describes the coupling strength between the QD and the leads. It is defined as $\Gamma={1\over2}(\Gamma^L+\Gamma^R)$ with $\Gamma^\alpha=2\pi |V_{\alpha}|^2\rho(\omega)$. We will ignore the
$\omega$-dependence of $\Gamma^\alpha$ since the electron density of
states in the leads, $\rho (\omega)$, can be usually viewed as
a constant. In
Eq.(\ref{transmission}) the retarded and advanced Green functions
in Fourier space are involved. These Green functions can be solved
by the equation-of-motion method. By a straightforward derivation,
we can obtain the matrix form of the retarded Green function.

\par
In this work, we would like to take our interest in the case of MZMs, i.e., $t_n=0$. The resulting retarded Green function is given by
\begin{eqnarray}
{\bf G}^r(\omega)=\left[\begin{array}{cccc} g_{e}(z)^{-1} &0&-\lambda_1& -\lambda_{2}\\
  0& g_{h}(z)^{-1} &\lambda_1^*& \lambda^*_{2}\\
  -\lambda^*_1&\lambda_1 & z &0\\
 -\lambda^*_{2} &\lambda_{2}&0&z
\end{array}\right]^{-1}\ \label{green},
\end{eqnarray}
with $z=\omega+i0^+$.
$g_{e(h)}(z)=[z\mp\varepsilon_{d}+i\Gamma]^{-1}$ is the
zero-order Green function of the QD unperturbed by the MBSs. In this work, we assume $\lambda_1=|\lambda_1|$ and $\lambda_2=i|\lambda_2|e^{i\varphi/2}$, where $\varphi$ is the phase difference between the two TS nanowires. And then, one can find that
\begin{eqnarray}
G^r_{dd}={1\over\cal D}[(\omega+\varepsilon_d+i\Gamma)\omega^2-\omega(|\lambda_1|^2+|\lambda_2|^2)]
\end{eqnarray}
with ${\cal D}=(\omega-\varepsilon_d+i\Gamma)(\omega+\varepsilon_d+i\Gamma)\omega^2-2\omega(\omega+\varepsilon_d+i\Gamma)(|\lambda_1|^2+|\lambda_2|^2)+2(1+\cos\varphi)|\lambda_1|^2|\lambda_2|^2$.
\par
In the linear response regime, we can expand the electric and heat
currents up to the linear terms of a temperature gradient $\delta
T=T_L-T_R$ to a thermoelectric voltage $\delta V=V_L-V_R$. The
transport coefficients $L_{ij}$ are defined by the relations
\begin{equation}
\left (
\begin{array}{ccc}
J^L_e\\
J^L_Q
\end{array}\right )=
\left (
\begin{array}{ccc}
L_{11}&L_{12}\\
L_{21}&L_{22}
\end{array}\right )
\left (
\begin{array}{ccc}
V_L-V_R\\
T_L-T_R
\end{array}\right ).\label{t11}
\end{equation}
and can be expressed in terms of the transport integral $K_n={1\over
h}\int d\omega (-{\partial f\over \partial
\omega})(\omega-\mu)^n\tau(\omega)$ as $L_{11}=e^2K_0$,
$L_{21}=L_{12}T=-eK_1$, and $L_{22}=K_2/T$. $\mu$ is the system's chemical potential. Then the linear response
conductance ${\cal {G}}=\lim_{V\rightarrow0}{dJ_e\over dV}=L_{11}$
is given by the equation
\begin{eqnarray}
{\cal {G}}=e^2K_0.
\end{eqnarray}
The thermopower of a QD system in a configuration with two normal leads can
be found in an open circuit by measuring the induced voltage drop
across a QD when the temperature difference is
applied between the two leads. The thermopower is defined by the relation
\begin{eqnarray}
S=-{\delta V\over \delta T}\mid_{J=0},
\end{eqnarray}
and can be expressed as
\begin{eqnarray}
S=-{1\over eT}{K_1\over K_0}.
\end{eqnarray}
The electronic contribution to the thermal conductance defined by
$\kappa_e={\Delta J_Q\over \Delta T}$ can be expressed by
\begin{eqnarray}
\kappa_e=K_1eS+{K_2\over T}.
\end{eqnarray}
Based on these formulas, the thermoelectric efficiency in such a structure can be evaluated, since its corresponding figure of merit is defined as
\begin{equation}
ZT=S^2{\cal G} T/\kappa_e,\label{ZT}
\end{equation}
in which the phonon-contributed thermal conductance has been ignored.

\begin{figure}
\centering \scalebox{0.27}{\includegraphics{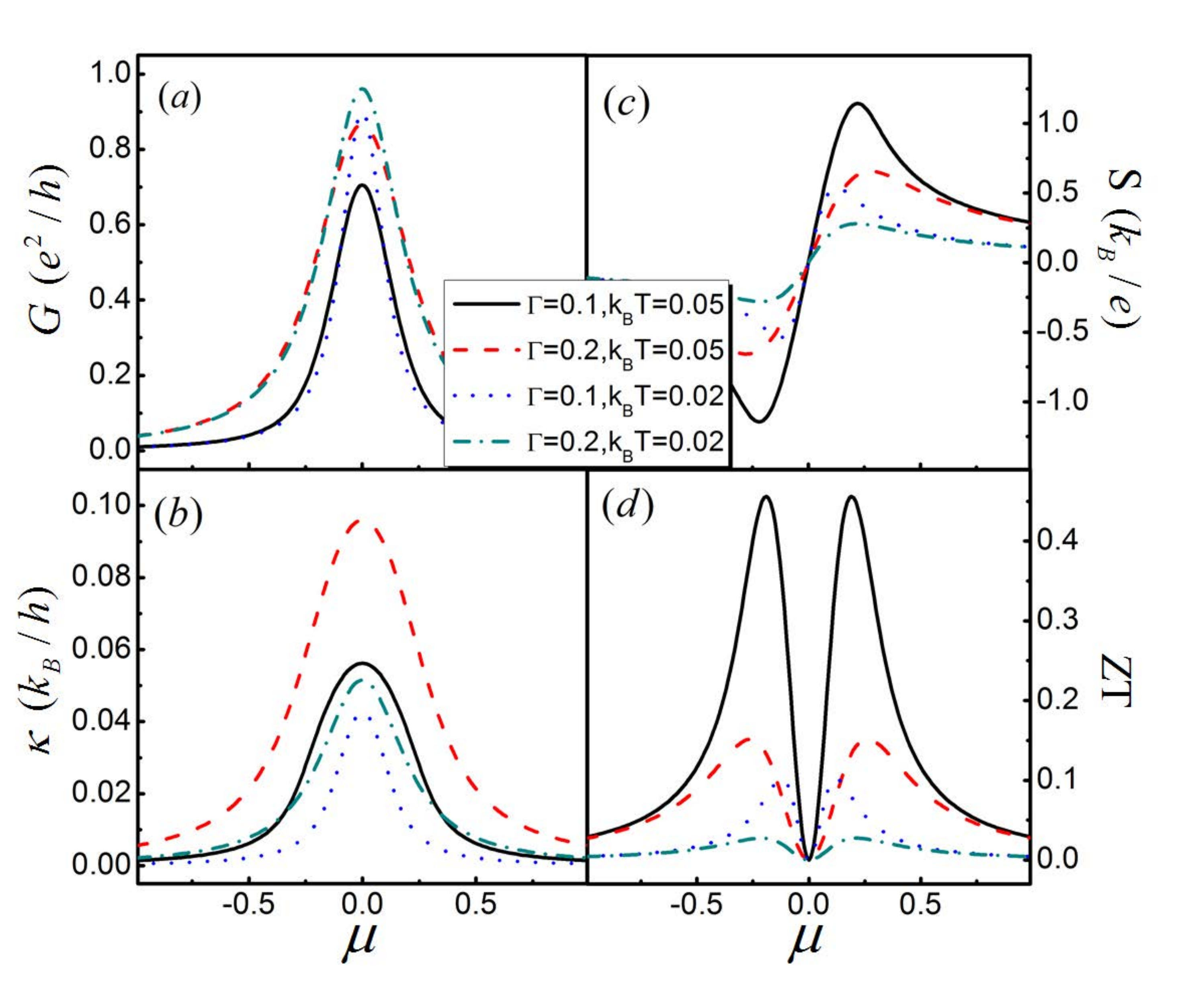}} \caption{
Thermoelectric quantities of one single-QD structure as functions of the system's chemical potential $\mu$, in the absence of QD-MZM couplings. (a)-(b) show the numerical results of the electronic conductance, thermal conductance, thermopover, and figure of merit. The relevant parameters are taken to be $\Gamma=0.1$ and $0.2$, $k_BT=0.02$ and $0.05$, respectively. }
\end{figure}

\section{Numerical results and discussions \label{result2}}
Following the formulations developed in the above section, we begin to perform the numerical calculation to investigate the thermoelectric properties of our considered QD structure with side-coupled MZMs. In the context, we are only interested in the case of symmetric QD-lead coupling, i.e., $\Gamma^L=\Gamma^R=\Gamma$.
\par
First of all, we ignore the QD-MZM coupling and present the thermoelectric properties of the single-QD structure. The numerical results are shown in Fig.2. In this figure, the QD-lead coupling is taken to be $\Gamma=0.1$ and $0.2$, and temperature is assumed to be $k_BT=0.02$ and $0.05$ for comparison. With respect to the QD level, we consider it to be zero. Fig.2(a)-(d) show the spectra of the electronic conductance, thermal conductance, thermopower, and figure of merit, respectively. Figure 2(a) shows that with the increase of the QD-lead coupling, the electronic conductance can be enhanced, while it can be suppressed by the increase of temperature. In Fig.2(c), it shows that the thermal conductance exhibits the similar property to electronic conductance. This should be attributed to the fact that the thermal conductance is mainly contributed by the electron motion.
As for the thermopower, in Fig.2(b) one can see that the Seebeck coefficient is the odd function of the chemical potential $\mu$, and it is positive (negative) when $\mu$ is greater (less) than zero. Besides, the Seebeck-coefficient magnitude decreases following the increase of the QD-lead coupling. However, it does not undergo the monotonous change with the increase of temperature. Namely, the maximum of the Seebeck coefficient in the case of $k_BT=0.05$ is greater than the case of $k_BT=0.02$. The thermoelectric efficiency is characterized by the extremum of the figure of merit, i.e., $ZT$. In Fig.2(d), we observe that similar to the thermopower, the thermoelectric efficiency in the case of $k_BT=0.05$ and $\Gamma=0.1$ is prior to the other cases, and the maximum of $ZT$ has an opportunity to be greater than 0.4.
\begin{figure}[htb]
\centering \scalebox{0.27}{\includegraphics{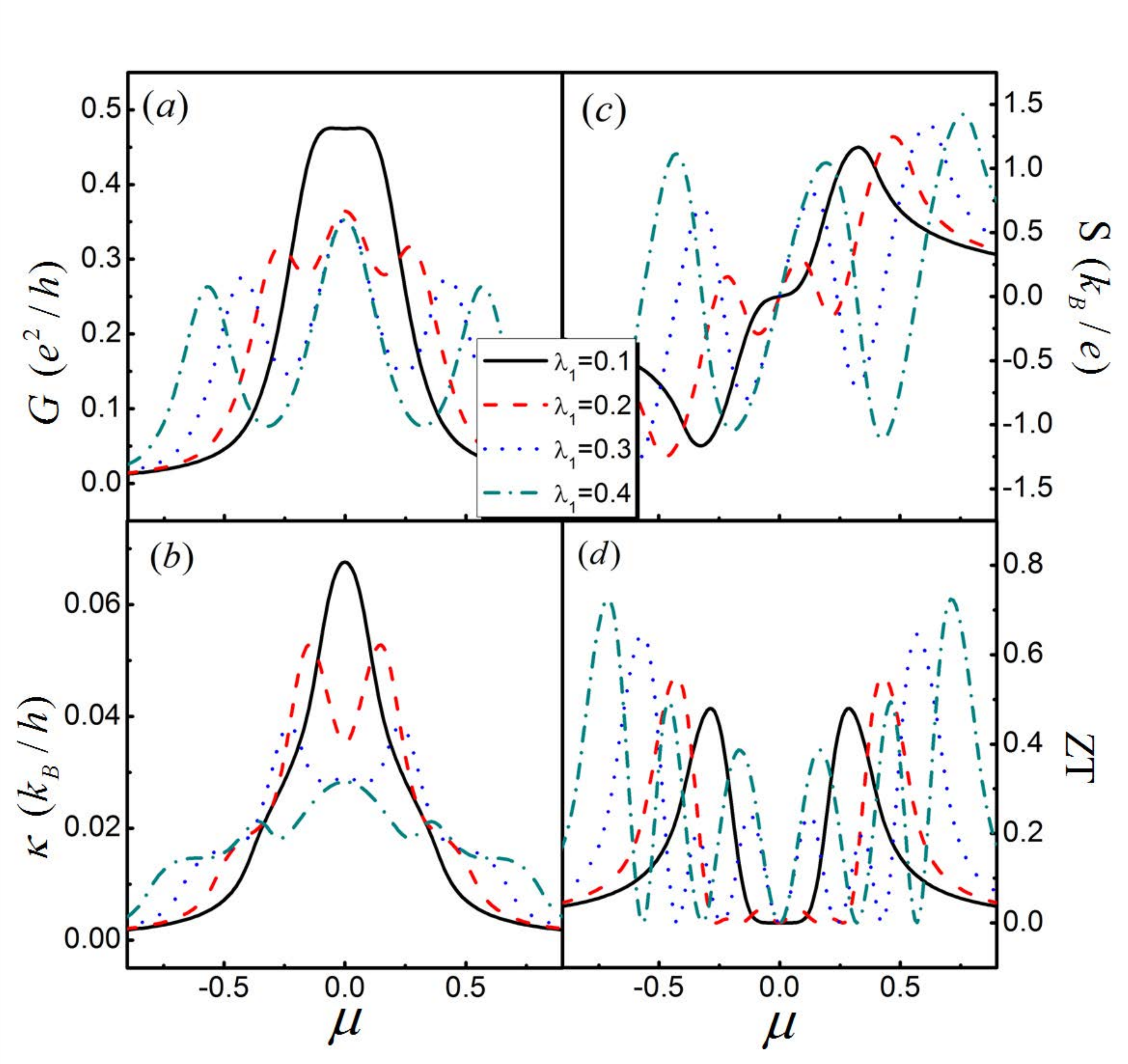}} \caption{
(a)-(d) Spectra of electronic conductance, thermal conductance, thermopover, and figure of merit in the single-QD structure, with one MZM coupled to the QD. The QD-lead couplings and temperature are taken to be $\Gamma=0.1$ and $k_BT=0.05$. }
\end{figure}
\par
Next, we begin to consider MZMs to couple to the QD and study their influences on the thermoelectric effect. As a typical case, we take $\Gamma=0.1$ and $k_{B}T=0.05$ in this part. The single-MZM result is shown in Fig.3. In Fig.3(a) we find that with the increase of QD-MZM coupling, the conductance peak is split into three, accompanied by the suppression of its magnitude. Such a result can be explained as follows. In the presence of finite QD-MZM coupling, the QD Hamiltonian in the Nambu representation can be written as
\begin{eqnarray}
H=\left[\begin{array}{cccc} \varepsilon_d &0& \lambda_1\\
  0& -\varepsilon_d &-\lambda_1^*\\
  \lambda^*_1&-\lambda_1 & 0\\
\end{array}\right]\ \label{green}.
\end{eqnarray}
It is evident that the QD's eigenlevels are $E_{\pm}=\pm\sqrt{\varepsilon_d^{2}+2\lambda_1^{2}}$ and $E_0=0$, for the real $\lambda_1$. Since the peaks of the electronic conductance spectrum are related to the QD's eigenlevels, one can readily understand the splitting of the conductance peaks when the QD-MZM coupling is taken into account. With the help of this analysis, we can also know the thermal conductance variation in this process, as shown in Fig.3(b). Next for the thermopower and figure of merit, in Fig.3(c)-(d) it shows that with the increase of QD-MZM coupling, their curves oscillate more seriously, accompanied by the increment of the magnitudes of them. However, such an increase is relatively weak. For instance, in the case of $\lambda_1=0.4$, the maximum of the Seebeck coefficient is only about equal to $S_{max}=1.5$, whereas $ZT_{max}\approx 0.7$. These results mean that the coupling of one MZM to the QD cannot efficiently enhance the thermoelectric effect, though its modification.

\begin{figure}[htb]
\centering \scalebox{0.28}{\includegraphics{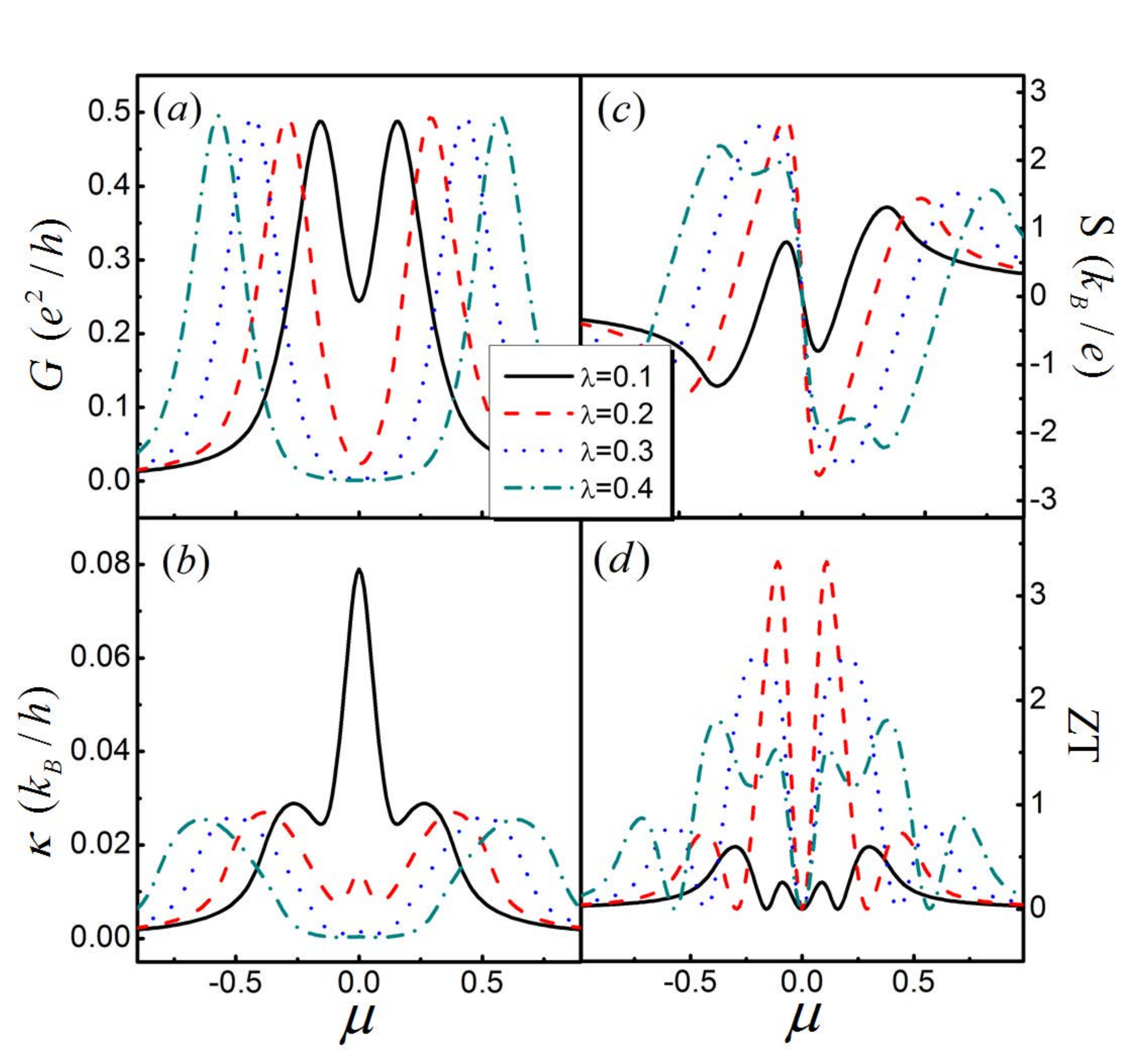}} \caption{
Thermoelectric quantities of the single-QD structure, in the presence of two MZMs coupled to the QD simultaneously. The QD-lead coupling and temperature are the same as those in Fig.3, respectively.      }
\end{figure}
\par
In the following, we continue to consider two MZMs to couple to the QD simultaneously, to discuss the change of the thermoelectric effect. The numerical results are shown in Fig.4. In this figure, we suppose $|\lambda_1|=|\lambda_2|=\lambda$ for simplicity. Firstly, from Fig.4(a), we can clearly find that differently from the above two cases, the electronic conductance spectrum exhibits two peaks, which are separated by the conductance around the point of $\mu=0$. The distance of them is proportional to the increase of $\lambda$. In order to clarify this result, we present the QD's Hamiltonian matrix, i.e.,
\begin{eqnarray}
H=\left[\begin{array}{cccc} \varepsilon_d &0& \lambda_1&\lambda_2\\
  0& -\varepsilon_d &-\lambda_1^*&-\lambda_2^*\\
  \lambda^*_1&-\lambda_1 & 0&0\\
  \lambda^*_2&-\lambda_2&0&0
\end{array}\right]\ .
\end{eqnarray}
One can then find that in the case of $|\lambda_1|=|\lambda_2|=\lambda$ and $\varphi=0$, the QD's eigenlevels are given by $E_{\pm}={1\over2}(\varepsilon_d \pm \sqrt{\varepsilon^2_d+8\lambda^2})$. Surely, the co-existence of two identical MZMs leads to the level degeneracy in this structure. For the typical case of $\varepsilon_d=0$, the eigenlevels will be simplified, i.e., $E_{\pm}=\pm\sqrt{2}\lambda$. Moreover in such a case, we are allowed to write out the expression of the transmission function $\tau(\omega)$, i.e.,
\begin{equation}
\tau(\omega)={\omega^2\Gamma^2\over(\omega^2-2\lambda^2)^2+\omega^2\Gamma^2}.
\end{equation}
It clearly shows that under the situation of $\omega=\pm\sqrt{2}\lambda$, the transmission function reaches unity. This exactly leads to the peaks in the electronic conductance spectrum. Meanwhile, one can readily find that at the low-energy limit, i.e., $\omega=0$, $\tau(\omega)$ will be equal to zero, which means the occurrence of the antiresonance phenomenon. This can be viewed as the reason of the conductance valley in Fig.4(a).
\par
For the other thermoelectric quantities, in Fig.4(b) we see that the thermal conductance is more sensitive to the increase of QD-MZM coupling, differently from the electronic conductance. It shows that when $\lambda=0.1$, the thermal conductance exhibits one peak at the position of $\mu=0$. With the increase of $\lambda$, this peak is suppressed apparently, and then the thermal conductance spectrum shows up as a two-peak structure. Next, Fig.4(c) shows that the increase of QD-MZM coupling does not only modify the oscillation manner of the Seebeck-coefficient curve, but also enhances the magnitude of it. For instance, in the case of $\lambda=0.2$, the magnitude of the Seebeck coefficient can reach 2.5, whereas the further increase of $\lambda$ cannot enhance the thermopower apparently. The thermopower increase certainly induces the enhancement of the thermoelectric efficiency. It can be readily seen in Fig.4(d) that the maximum of $ZT$ has an opportunity to arrive at 3.5, when $\lambda=0.2$. Therefore, we know that when two MZMs couple to one QD with the same strength, the thermoelectric effect will be enhanced to a great extent. It is certain that such a phenomenon originates from the occurrence of the antiresonance effect in the electron motion process, in the presence of two MZMs coupled to the QD.
\par
\begin{figure}
\centering \scalebox{0.18}{\includegraphics{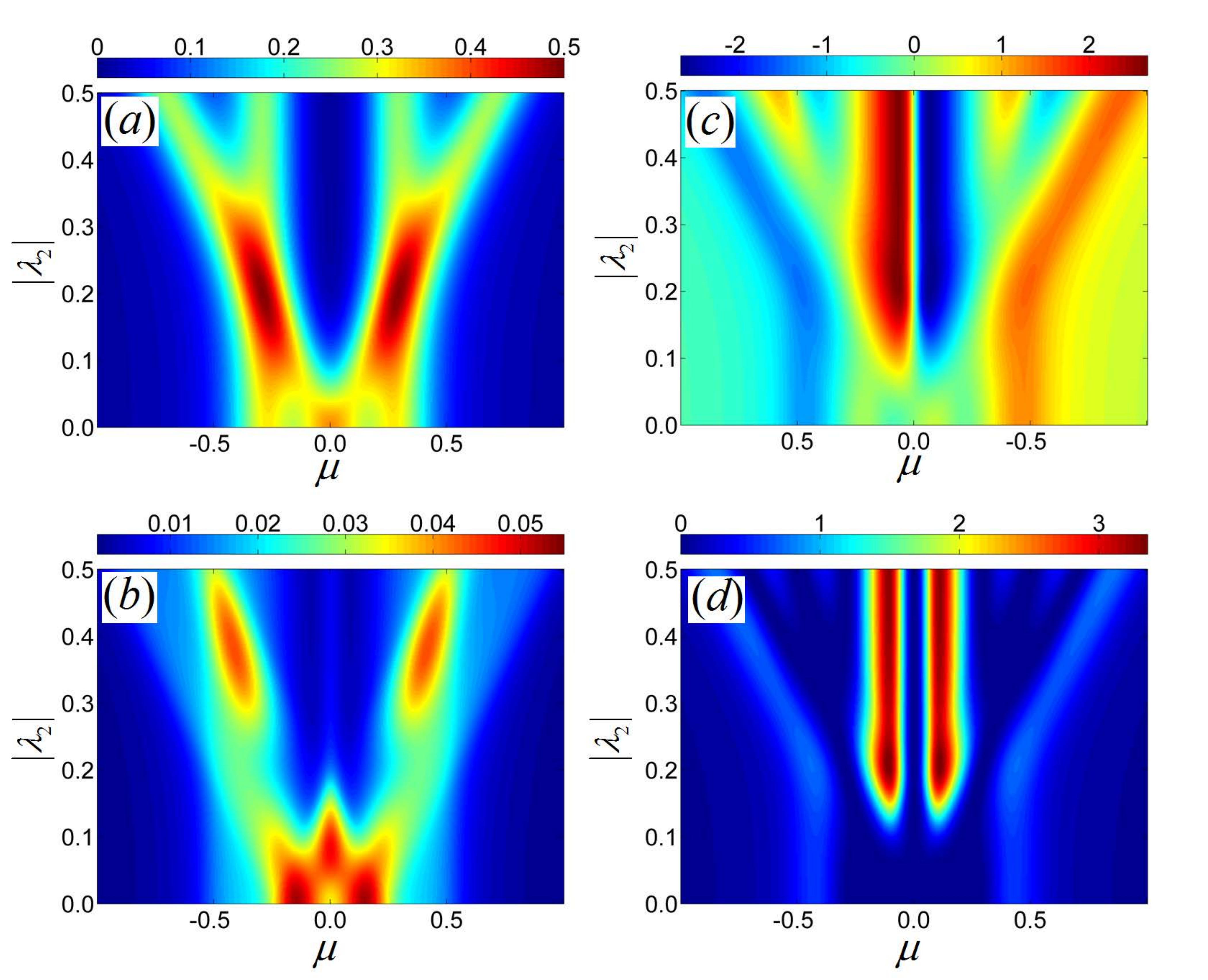}} \caption{
Results of thermoelectric quantities modulated by the difference between the two kinds of QD-MZM couplings. (a)-(d) Electronic conductance, thermal conductance, thermopover, and figure of merit.    }
\end{figure}
Encouraged by the notable thermoelectric result in Fig.4, we next would like to adjust the structural parameters to further clarify the thermoelectric effect. To begin with, in Fig.5 we investigate the influence of the difference between $\lambda_1$ and $\lambda_2$ on the thermoelectric quantity properties. For calculation, we take $\lambda_1=0.2$ and increase the value of $\lambda_2$. In Fig.5(a), we can find that in the increase of $|\lambda_2|$ the electronic conductance spectrum undergoes its three-step variations. Firstly, in the case of $|\lambda_2|<0.1$, three peaks appear in the conductance spectrum. Then with the following increase of $|\lambda_2|$, two peaks exist in the conductance spectrum, followed by their enhanced magnitudes. Such a phenomenon ends until $|\lambda_2|$ increases to $0.4$, where each conductance peak splits into two and their magnitudes are suppressed. In comparison with the electronic conductance, the maximal thermal conductance occurs in the region of $|\lambda_2|<0.1$, whereas it is weakened in the region of $0.1<|\lambda_2|<0.3$, as shown in Fig.5(b). This certainly leads to the enhancement of the thermoelectric effect, because the out-of-phase changes of the electronic and thermal conductances. Such a conclusion can be well verified by the results in Fig.5(c)-(d). As shown in Fig.5(c), when $|\lambda_2|>0.1$, the magnitude of the Seebeck efficient increases rapidly, about equal to $3.0$. However, its value seems to be independent of the further increase of $|\lambda_2|$. Similarly, the result in Fig.5(d) shows that the thermoelectric efficiency can be enhanced in this process, and the value of $ZT$ is allowed to get close to 3.5. From the results in Fig.5, we can find that the enhancement of thermoelectric effect does not depend on the identical QD-MZM couplings.

\begin{figure}[htb]
\centering \scalebox{0.18}{\includegraphics{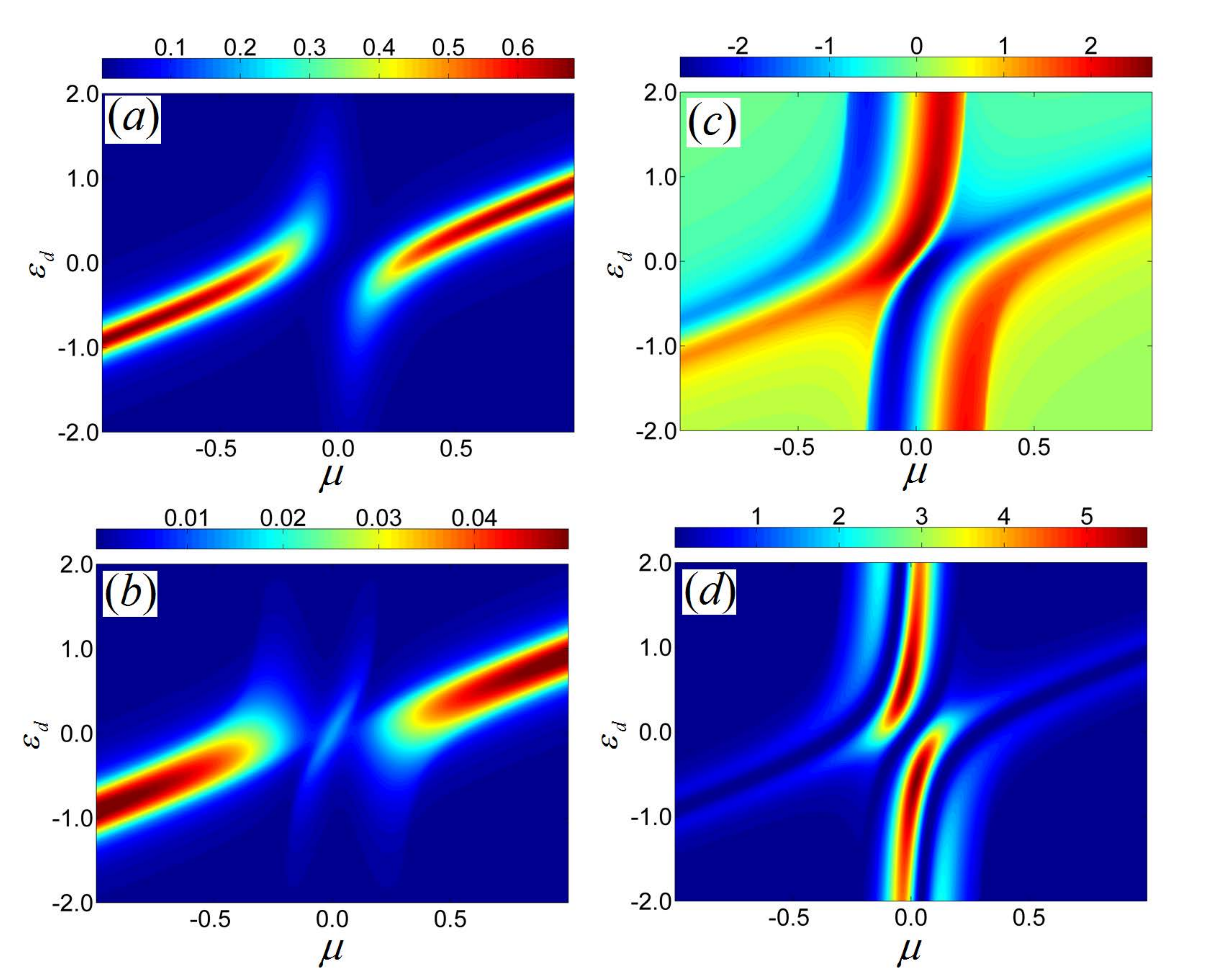}} \caption{
Spectra of thermoelectric parameters influenced by the shift of QD level $\varepsilon_d$. The other structural parameters are the same as Fig.4. }
\end{figure}
\par
Since the QD level can be shifted in experiment, we next would like to present the influence of the shift of the QD level on the thermoelectric effect. The main results are shown in Fig.6. Firstly, from Fig.6(a)-(b) we can see that both the spectra of the electronic and thermal conductances are asymmetric about the chemical potential $\mu$ and the QD level $\varepsilon_d$. When the chemical potential and QD level are adjusted simultaneously in the positive-energy (negative-energy) region, the conductance magnitudes can be increased, especially in the case of $\mu=\varepsilon_d$. Also note that in the low-energy region where $\varepsilon_d$ and $\mu$ get close to zero, the conductances are weakened. This should be attributed to the occurrence of the antiresonance effect.  Next in Fig.6(c), one can readily find that the thermopower can be increased by shifting the QD level away from the energy-zero point. Namely, in the case of $|\varepsilon_d|\approx 0.5$, the Seebeck coefficient arrives at its extremum with $|S|_{max}\approx 3.0$. And similar result can be observed in Fig.6(d). In this figure we find that when the QD level is shifted in the vicinity of $\varepsilon_d=0.5$, the value of $ZT$ have opportunities to reach its maximum with $ZT_{max}\approx5.5$. Up to now, we can ascertain that shifting the QD level is an efficient way to enhance the thermoelectric effect.
\par
In Fig.7, we are concerned about the impact of phase difference between the two TSs that yield MZMs. The other parameters are chosen to be $|\lambda_1|=|\lambda_2|=\lambda=0.2$ and $\varepsilon_d=0$. Firstly, in Fig.7(a), one can observe that with the increase of the superconducting phase difference, the magnitude of the electronic conductance first decreases and then increases. In such a process, the conductance peaks split in the complicated way. When $\varphi$ increases from zero to $\pi$, both of the two conductance peaks split into two. The two conductance peaks beside the point of $\mu=0$ get close to each other, and then they encounter in the case of $\varphi=\pi$. As a result, in this case, three peaks appear in the electronic conductance spectrum. Next, as the superconducting phase difference further increases, the opposite change manner takes place. Since the electronic conductance is related to the QD's eigenlevels in the Nambu representation, we would like to present the expression of them, i.e., $E=\pm\lambda\sqrt{2\mp 2\sin\varphi}$ under the situation of $\varepsilon_d=0$. It can found that in the general case, there are four different values for $E$. However, for $\varphi=2n\pi$, they are degenerated, with $E_{\pm}=\pm\sqrt{2}\lambda$. Alternatively, when $\varphi=(2n+1)\pi$, $E$ will present three values, i.e., $E_{\pm}=\pm 2 \lambda$ and $E_{0}=0$. Based on these results, one can well understand the change of the electronic conductance spectrum. Next, in Fig.7(b) it shows that the thermal conductance spectrum always exhibit three peaks, with the increase of the phase difference. What is notable is that when the TS phase difference departs from the value of $\varphi=2\pi$ (or $\varphi=0$), the thermal conductance is increased, completely from the change manner of the electronic conductance. Such a fact inevitably weakens the thermoelectric effect, according to Eq.(\ref{ZT}). As a consequence, we see in Fig.7(c)-(d) that only in the case of $\varphi=0$ (or $2\pi$), the thermoelectric effect is relatively active. Therefore, the TS phase difference is a negative factor for the enhancement of the thermoelectric effect.
\begin{figure}
\centering \scalebox{0.18}{\includegraphics{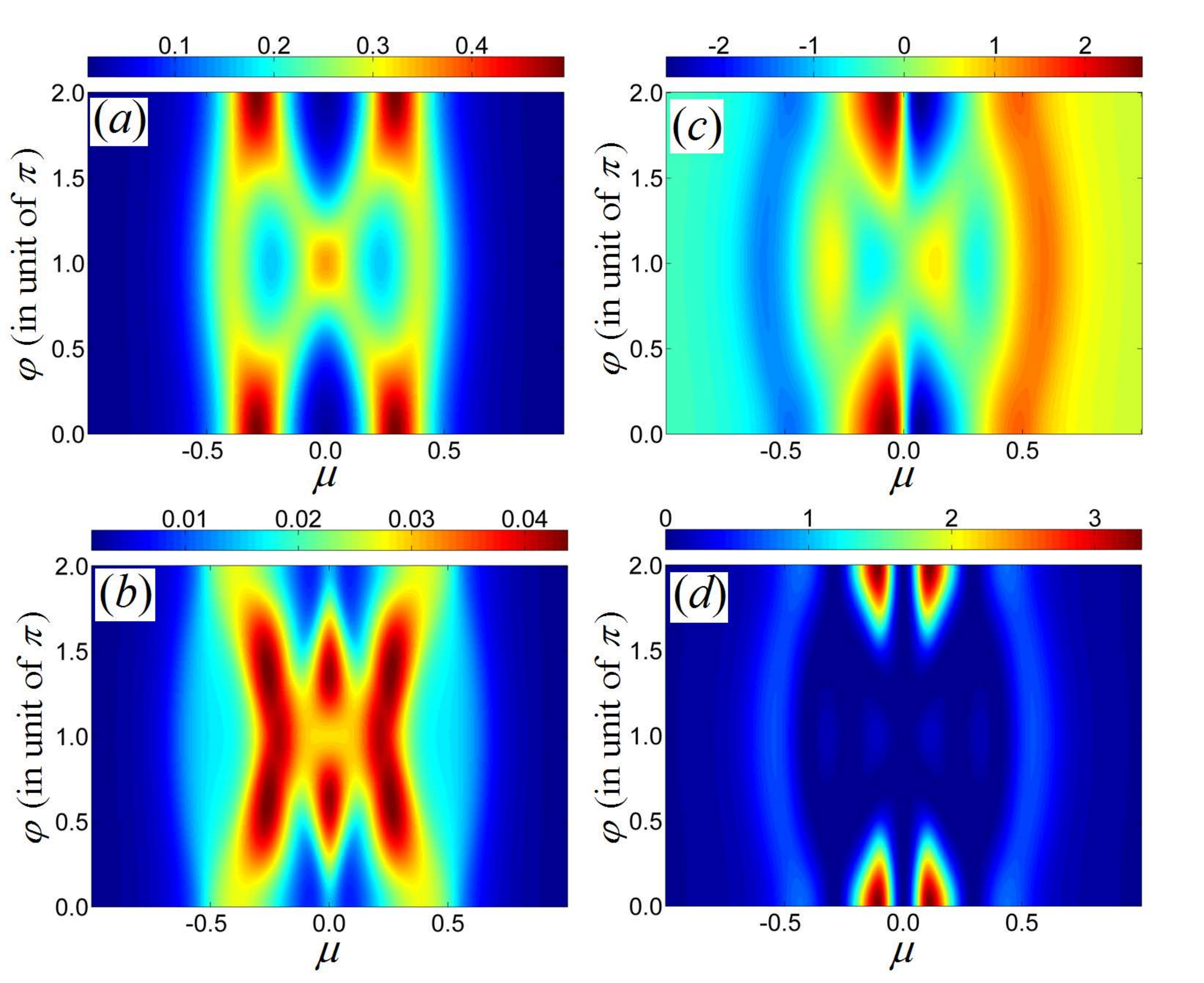}} \caption{
Influence of the change of TS phase difference on the thermoelectric effect of the single-QD structure with two side-coupled MZMs. }
\end{figure}

\par
In addition to the discussion above, we would like to make a remark regarding the many-body
effect which we have by far ignored. However, the many-body effect
is an important origin for the peculiar transport properties in the QD.
Therefore, it is supposed to influence the thermoelectric effect to some
extent. Usually, the many-body effect is incorporated by considering
only the intradot Coulomb repulsion, i.e., the Hubbard term. If the
Hubbard interaction is not very strong, we can truncate the
equations of motion of the Green functions to the second
order. By a straightforward derivation, we find
that the theoretical descriptions developed
above are still valid. But the Green
functions $g_{e(h)\sigma}$ should be redefined as
\begin{equation}
g_{e(h)\sigma}(z)=\big[\frac{z\mp\varepsilon_d}{1+\frac{U\langle
n_{\bar{\sigma}}\rangle}{z\mp\varepsilon_d\mp U}}+i\Gamma\big]^{-1}.
\end{equation}
Albeit this change, we can anticipate that the thermoelectric effect in the noninteracting case will hold. The reason can be explained as follows. Firstly, such an approximation is only to induce the splitting of the QD level from $\varepsilon_d$ to $\varepsilon_d$ and $\varepsilon_d+U$ ($U$ is the Coulomb strength), but does not lead to the appearance of new physics picture\cite{BB,Dicke}. Secondly, due to the helix of the MZMs, we can consider that they only couple to one spin state of the QD, whereas, the other spin state decouples from the MZMs. This further weakens the effect of the Coulomb interaction\cite{CC1,CC2,gong,CC3}. Therefore, we can believe that the Coulomb term does not result in the substantial change of the noninteracting thermoelectric properties. When
the electron interaction is very strong, one need extend the
theoretical treatment beyond this approximation. Then the further modification to the
thermoelectric effect will naturally arise. This interesting subject will be left for our future study.
\section{summary\label{summary}}
To summarize, in this work we have investigated the thermoelectric effect in one QD system, by coupling two MZMs laterally to the QD, independently. As a result, it has been found that if only one MZM couples to the QD, the thermoelectric effect cannot be strengthened efficiently. However, the co-existence of two MZMs can enhance the thermoelectric effect to a great extent. Namely, the thermoelectric efficiency, i.e, $ZT$ has an opportunity to reach 5.5. Via presenting the parameter influences, this result is analyzed in detail. We believe that the results in this work can be helpful in understanding the quantum transport property of the MZM.

\section*{Acknowledgements}
This work was financially supported by the National Natural
Science Foundation of China (Grant No. 11604221), the Natural Science
Foundation of Liaoning province of China (Grant No. 20170540665), and the Fundamental Research
Funds for the Central Universities (Grant No. N160504009).

\clearpage

\bigskip

\end{document}